\documentclass[nofootinbib,showpacs,preprintnumbers,amsmath,amssymb, prd]{revtex4}
\usepackage{color}
\usepackage{amsmath,amssymb,graphicx}
\usepackage{epsf}
\usepackage{epstopdf}
\usepackage {amssymb}
\newcommand{\nc}{\newcommand}
\nc{\ba}{\begin{eqnarray}}
\nc{\ea}{\end{eqnarray}}
\newcommand\be{\begin{equation}}
\newcommand\ee{\end{equation}}

\newcommand\mPl{{M_{\rm P}}}

\newcommand{\calR}{{\cal{R}}}

\nc{\x}{{\bf{x}}}

\nc{\pp}{{\bf{p}}}
\nc{\px}{P_{,X}}
\nc{\pxt}{P_{,X \phi}}
\nc{\pxx}{P_{,XX}}
\nc{\pxxx}{P_{,XXX}}
\nc{\pt}{P_{,\phi}}
\nc{\ptt}{P_{,\phi \phi}}
\nc{\pttt}{P_{,\phi \phi \phi }}
\nc{\pxtt}{P_{,X \phi \phi}}
\nc{\pxxt}{P_{,X X \phi}}

\newcommand{\bea}{\begin{eqnarray}}
\newcommand{\eea}{\end{eqnarray}}

\def\bk{{\bf k}}

\def\bx{{\bf x}}

\def\CL{{\cal L}}

\def\half{\frac{1}{2}}

\newcommand{\calP}{{\cal P}}

\begin{document}
\title{In-in and $\delta N$ calculations of the bispectrum from
non-attractor single-field inflation}

\author{Xingang Chen$^{1,2}$}
\author{Hassan Firouzjahi$^{3}$}
\author{Eiichiro Komatsu$^{4,5,6}$}
\author{Mohammad Hossein Namjoo$^{7}$}
\author{Misao Sasaki$^{8}$}

\affiliation{$^1$Centre for Theoretical Cosmology, DAMTP, University of Cambridge, Cambridge CB3 0WA, UK}
\affiliation{$^2$Department of Physics, The University of Texas at Dallas, Richardson, TX 75083, USA}
\affiliation{$^3$School of Astronomy, Institute for Research in
Fundamental Sciences (IPM),
P. O. Box 19395-5531,
Tehran, Iran}
\affiliation{$^4$Max-Planck-Institut f\"{u}r Astrophysik,
Karl-Schwarzschild Str. 1, 85741 Garching, Germany}
\affiliation{$^5$Kavli Institute for the Physics and
Mathematics of the Universe, Todai Institutes for Advanced Study, the
University of Tokyo, Kashiwa, Japan 277-8583 (Kavli IPMU, WPI)}
\affiliation{$^6$Texas Cosmology Center and the Department of Astronomy,
The University of Texas at Austin, 1 University Station, C1400, Austin,
TX 78712, USA}
\affiliation{$^7$School of Physics, Institute for Research in
Fundamental Sciences (IPM),
P. O. Box 19395-5531,
Tehran, Iran}
\affiliation{$^8$Yukawa Institute for theoretical Physics,
 Kyoto University, Kyoto 606-8502, Japan}

\preprint{YITP-13-81}

\begin{abstract}
 In non-attractor single-field inflation models producing
 a scale-invariant power spectrum, the curvature perturbation on
 super-horizon scales grows as ${\cal R}\propto a^3$. This is so far the
 only known class of
 self-consistent single-field models with a
 Bunch-Davies initial state that can produce a large squeezed-limit
 bispectrum violating Maldacena's consistency relation. Given the
 importance of this result, we calculate the bispectrum with three
 different methods: using quantum field theory calculations in two
 different gauges, and classical calculations (the $\delta N$
 formalism). All the results agree, giving the local-form bispectrum
 parameter of $f^{\rm
 local}_{NL}=5(1+c_s^2)/(4c_s^2)$. This result is valid for arbitrary
 values of the speed of sound parameter, $c_s$, for a particular
 non-attractor model we consider in this paper.
\end{abstract}

\maketitle

%%%%%%%%%%%%%%%%%%%%%%%%%%%%%%%%%%%%%%%%%%%%%%%%%%
\section{Introduction}
Recent high-precision measurements of fluctuations in the
cosmic microwave background (CMB) from WMAP
\cite{Bennett:2012zja,Hinshaw:2012fq} and Planck \cite{Ade:2013lta,
Ade:2013uln} strongly support inflation \cite{Starobinsky:1980te,Sato:1980yn,Guth:1980zm,Linde:1981mu,Albrecht:1982wi} as the
leading theory of the early universe
and structure formation. The simplest models of inflation are based on a
single scalar field slowly rolling down on a flat potential. These models
predict adiabatic, almost scale-invariant, and almost Gaussian
primordial fluctuations in the CMB
\cite{Mukhanov:1981xt,Guth:1982ec,Starobinsky:1982ee,Hawking:1982cz,Bardeen:1983qw}, in agreement with all the data we have today.

Non-Gaussianity has emerged as a powerful observational tool to
discriminate between different inflationary models (see
\cite{Bartolo:2004if,Komatsu:2009kd,Chen:2010xka,Komatsu:2010hc} for
reviews).
While the Planck data show no evidence for non-Gaussianity,
the current limits \cite{Ade:2013ydc} are yet to reach the levels
expected from the
simple single-field models.\footnote{By
``simple single-field models,'' we refer to single-field models with the
canonical kinetic term and a Bunch-Davies initial state, which have
approached attractor solutions. Namely, one of the two solutions of the
curvature perturbation on super-horizon scales is a constant, and the
other is a decaying solution. In these conditions, the condition of the canonical kinetic term can be generalized to non-canonical terms without changing the consistency condition.} Therefore, there is still
significant room for non-Gaussianity to be found, with the amplitude
much greater than that predicted by the simple models.

In particular, the so-called {\it squeezed-limit} of the
bispectrum of primordial curvature perturbations, ${\cal R}$, plays a special
role. The bispectrum is defined by $\langle \calR_{\bk_1} \calR_{\bk_2}
\calR_{\bk_3} \rangle=(2\pi)^3\delta^3(\sum_i \bk_i) B_{\cal R}(k_1,k_2,k_3)$,
and the squeezed limit is the limit in which one of the wavenumbers is
much smaller than the other two, i.e., $k_3\ll k_1\approx k_2$. One can
show that all of the simple single-field models$^1$ satisfy Maldacena's
consistency relation \cite{Maldacena:2002vr} given by
\ba
\label{consistency_cond}
B_{\cal R}(k_1,k_2,k_3) \to
(1-n_s) \frac{(2\pi)^4}{4k_1^3 k_3^3} \calP_\calR(k_1) \calP_\calR(k_3)~,
\ea
for $k_3 \ll k_1\approx k_2$.
Here, the curvature-perturbation power spectrum per logarithmic intervals
in momentum space, ${\cal P}_{\cal R}(k)\propto k^{n_s-1}$, is defined by
$\langle \calR_{\bk} \calR_{\bk'}\rangle = (2\pi)^3
\delta^3(\bk+\bk')\frac{2\pi^2}{k^3}\calP_\calR(k)$.
A convenient quantity to express the magnitude of the
bispectrum in the squeezed limit is the (local form) $f_{NL}$ parameter
defined by \cite{Komatsu:2001rj}
\begin{equation}
 \frac65 f_{NL}^{\mathrm{local}} \equiv
\frac{B_{\cal R}(k_1,k_2,k_3)}{\frac{(2\pi)^4}{4k_1^3 k_3^3}
\calP_\calR(k_1) \calP_\calR(k_3)+{\mbox{2 perm.}}}~,
\end{equation}
which approaches  $\frac65f_{NL}^{\mathrm{local}}\to \frac12(1-n_s)$ for $k_3
\ll k_1\approx k_2$.

Usually, single-field inflationary models predict an almost scale-invariant
spectrum, i.e., $1-n_s ={\cal{O}} (\epsilon, \eta)$, where $\epsilon$
and $\eta$ are the slow-roll parameters and they are of
order ${\cal O}(10^{-2})$ or smaller. As a result,  Maldacena's
analysis \cite{Maldacena:2002vr} shows that
all of the simple single-field models$^1$ give
$f_{NL}^{\mathrm{local}} =\frac5{12}(1- n_s) ={\cal{O}} (10^{-2})$.
An intuitive way to understand Maldacena's consistency relation is to
note that the large-scale mode, $k_3$, leaving the horizon long before
the small-scale modes,  $k_1$ and $k_2$, provides a constant re-scaling
of the background scale factor (hence the comoving
coordinates) for the small-scale modes
\cite{Creminelli:2004yq,Ganc:2010ff}.

Until recently, the only known class of single-field
inflation models which violate
Maldacena's consistency relation (for finite values of $k_3$) were the models
 with non-Bunch-Davies initial states
\cite{Chen:2010xka, Agullo:2010ws, Ganc:2011dy, Chialva:2011hc, Creminelli:2011rh,
Ganc:2012ae,
Agullo:2012cs, Agarwal:2012mq, Ashoorioon:2010xg} (also see
\cite{Chen:2006nt,Holman:2007na, Meerburg:2009ys} for earlier work
studying the effects of non-Bunch-Davies initial states).

However, Refs.~\cite{Namjoo:2012aa,
Chen:2013aj,Martin:2012pe,Huang:2013lda, Chen:2013kta} find that
single-field inflation models containing a non-attractor
phase at the initial stage of inflation can yield
$f_{NL}^{\rm local}$ that violates the consistency relation, without
invoking non-Bunch-Davies initial states. (Also see
\cite{Kinney:2005vj} for earlier work on non-attractor inflation
models.) In conventional models of single-field inflation,
one of the two solutions of the curvature perturbation on super-horizon
scales remains constant, and the other solution decays. On the other hand,
in non-attractor models of inflation,
what-would-be a decaying mode of the curvature perturbation in
conventional models of single-field inflation {\it grows} and
dominates over the constant mode during the non-attractor phase of
inflation.
This time evolution of the curvature perturbation violates
the consistency relation, as one can not simply absorb the effects of
the long-wavelength mode, ${\cal R}_{k_3}$, into a constant re-scaling of the
background scale factor for the short-wavelength modes. This
property thus calls for explicit calculations.

In the non-attractor models explored so far, one of the
slow-roll parameters decays as $\epsilon\equiv
\dot{\phi}^2/(2H^2)\propto 1/a^6$ and the curvature
perturbation on super-horizon scales grows as ${\cal R}\propto a^3$
during the non-attractor phase.
The simplest example is given by a scalar field with the canonical kinetic
term rolling on a {\it constant} potential \cite{Namjoo:2012aa}. The
kinetic energy of the scalar field is thus given by the initial
velocity, which decays as $\dot{\phi}^2\propto a^{-6}$. This gives
$\epsilon\propto 1/a^6$, and ${\cal R}=-H/\sqrt{2\epsilon}\propto
a^3$. Using both the quantum field theory calculation in the comoving
gauge \cite{Maldacena:2002vr} and the $\delta N$ formalism
\cite{Starobinsky:1982ee,Sasaki:1995aw, Sasaki:1998ug, Wands:2000dp,
Lyth:2004gb, Lyth:2005fi}, Ref.~\cite{Namjoo:2012aa} finds $f_{NL}^{\rm
local}=5/2\gg 1-n_s$, violating the consistency relation (also see
\cite{Martin:2012pe}).
The second example has a scalar field with a non-canonical kinetic term,
yielding the speed of sound of $c_s\ll 1$ \cite{Chen:2013aj} (see
Section~\ref{model-background} for the details of this set up). The
non-attractor inflation is still driven by a constant potential, leading
to $\epsilon\propto 1/a^6$ and ${\cal R}\propto a^3$.
Using  the quantum field theory calculation in the comoving gauge as
well as in the flat gauge \cite{Seery:2005wm}, Ref.~\cite{Chen:2013aj}
finds $f_{NL}^{\rm local}\simeq 5/(4c_s^2)\gg 1-n_s$ (for $c_s\ll 1$),
again violating the consistency relation (also see \cite{Huang:2013lda}).

In this paper, we first provide more detailed derivations of the
quantum field theory calculations used by Ref.~\cite{Chen:2013aj}.
We first show that the in-in formalism in
the comoving gauge yields $f_{NL}^{\rm local}=5(1+c_s^2)/(4c_s^2)$ for
arbitrary values of $c_s$.
We then present a new derivation of the same result using the $\delta N$
formalism. This is a non-trivial task: the usual application of the
$\delta N$ formalism is limited to the case in which attractor
solutions have been reached. Then, one needs to consider
derivatives of the number of $e$-folds of inflation, $N$, only with respect to the value of a scalar field, $\phi$, on the initial flat hypersurface. However,
one must take into account the full phase space, i.e., the values of
{\it both} $\phi$ and $\dot{\phi}$, when attractor solutions have not yet been
reached \cite{Namjoo:2012aa}.

One may worry about validity of the classical calculation
such as the $\delta N$ formalism when $c_s\ll 1$, as the $\delta N$
formalism is based on the gradient expansion~\cite{Salopek:1990jq}
 and thus ignores non-Gaussian contributions from modes at the horizon
 crossing \cite{Lyth:2004gb,Sugiyama:2012tj}. 
However, as we shall show below, the dynamics
responsible for the interactions between the modes and the generation of
the local-type non-Gaussianity in our models happens on super-horizon scales.
The $\delta N$ formalism thus gives accurate results, which are
insensitive to intrinsic non-Gaussianities generated at the horizon
crossing.

The rest of the paper is organized as follows. In Section
\ref{model-background} we present our set up. In Section
\ref{linear-perturbation} we present the linear cosmological
perturbation for our model and calculate the power
spectrum. In Section \ref{in-in} we calculate the
bispectrum using the in-in formalism in both the comoving and flat
gauges. In Section \ref{delta-N} we calculate the bispectrum  using the
$\delta N $ formalism. We conclude in Section
\ref{conclusion}. In the Appendix we show that the actions in the
flat and comoving gauges are equivalent to each other at
the leading order in $c_s^{-2}\gg 1$.
%%%%%%%%%%%%%%%%%%%%%%%%%%%%%%%%%%%%%%%%%%%%%%%%%%

\section{Non-attractor background with large self-interactions}
\label{model-background}

Here we present our set up. The model is the same as that studied in
Ref.~\cite{ Chen:2013aj} with the following action for a scalar field
with a non-standard kinetic energy such as in models of k-inflation
\cite{ArmendarizPicon:1999rj}:
\ba
S = \int  dt \, d^3 x \, P(X, \phi)~,
\ea
where $X \equiv -\frac{1}{2} \partial_\mu \phi \partial^\mu \phi $, and
\ba
\label{model}
 P(X,\phi) = X+ \dfrac{ X^\alpha }{M^{4\alpha - 4}}  -V(\phi) \qquad , \qquad V(\phi)= V_0 + v \left( \dfrac{ \phi }{\mPl} \right)^\beta~,
\ea
with $M$, $\alpha$, $v$, $V_0$, and $\beta$ being free constant
parameters.  As studied in Ref.~\cite{Chen:2013aj}, inflation has two
phases: the non-attractor phase, followed by the attractor phase.

As we shall show below, the second term of the potential is
negative, $v<0$. Inflation during the non-attractor phase is driven by a
constant term in the potential, $V_0$. The initial velocity of $\phi$ is
arranged
such that the field climbs up the potential initially,
and the first term of $P(X, \phi)$, i.e., the term linear in
$X$, is sub-dominant during the non-attractor phase. Towards the end of
the non-attractor phase (at which the kinetic energy decays
sufficiently), the term linear in $X$ dominates and the second phase of
inflation starts. We assume that the second phase is a usual slow-roll
inflation in which the super-horizon modes are frozen. However, the
crucial point is that the curvature perturbation is not conserved on
super-horizon scales during the early non-attractor phase. We find a
large non-Gaussianity when the CMB modes leave the horizon during the
non-attractor phase.

The background equations of motion are
\ba
3 \mPl^2 H^2 &=& 2   X P_{,X}-  P \, ,
\\
\mPl^2 \dot H &=& - X P_{,X} \, ,
\ea
and
\ba
\label{KG-eq0}
\left(P_{,X} + 2 X  P_{,XX} \right) \, \ddot \phi+ 3 H  \dot \phi P_{,X}  + 2 X \pxt-\pt =0 \, ,
\ea
where a dot indicates the derivative with respect to the cosmic time,
$t$, and $H\equiv \dot a/a$
is the Hubble constant during inflation. As usual we use the convention that
$P_{, X} \equiv \partial P/\partial X$ and so on.

Here we define the sound speed, $c_s$, and slow-roll parameters,
$\epsilon$ and $\eta$, by
\ba
c_s^2 \equiv \dfrac{\px}{\px+2 X \pxx}\, ,
\hspace{3cm}
\ea
\ba
\epsilon &\equiv& - \dfrac{\dot H}{H^2} = \dfrac{ X \px}{\mPl^2 H^2} \, ,
\\
\eta &\equiv& \dfrac{\dot \epsilon}{H \epsilon} = \dfrac{\ddot \phi}{H
\dot \phi} \left ( 1+ \dfrac{1}{c_s^2}\right) + \dfrac{\dot \phi \pxt
}{H \px} +2 \epsilon \, .
\ea
For future references, we also define the following variables:
\ba
\Sigma &\equiv& X P_{,X} + 2 X^2 P_{,XX}=\dfrac{H^2 \mPl^2 \epsilon}{c_s^2} \, ,
\\
\lambda &\equiv& X^2 P_{,XX} + \dfrac{2}{3} X^3 P_{,XXX} \, .
\label{lambda}
\ea

To support inflation we
assume that the constant potential term
dominates in the total energy density such that
\ba
3 \mPl^2 H^2 \simeq V_0\, .
\ea
Let us first consider the non-attractor phase in which the term linear in
$X$ can be neglected. To be able to do analytic calculation, we require
$c_s$ and $\eta$ to be nearly constant. The latter requirement
implies $\epsilon\propto a^\eta$. We will check below that
our Lagrangian can satisfy these requirements. The sound speed during
the non-attractor phase is given by
\ba
\label{sound-speed}
c_s^2 \simeq \dfrac{1}{2\alpha - 1} \, .
\ea

Since $P_{,X\phi}=0$, the Klein-Gordon equation Eq. (\ref{KG-eq0}) can be rewritten as
\ba
\label{KG-eq}
\dfrac{P_{,X}}{c_s^2}  \dot X + 6 H  X P_{,X}- P_{,\phi} \dot \phi =0 \, .
\ea
Finding analytical solutions of the above equation is not easy. Instead,
we propose the following ansatz:
\ba
\label{ansatz}
\phi(t) = {\rm const}\cdot a^\kappa  \, ,
\ea
where $\kappa$ is a constant and should be determined by the consistency of the equations. By this ansatz, and noting that $H$ is nearly constant, we have
\ba
\label{sol1}
\dot \phi \simeq H \kappa \phi \quad , \quad  \ddot  \phi \simeq H^2 \kappa^2 \phi \quad , \quad  X   \simeq \dfrac{1}{2} H^2 \kappa^2 \phi^2 \, .
\ea
Using the above relations we obtain two equations to solve
Eq. \eqref{KG-eq}, one for the cancellation of powers of $\phi$ and the
other for the cancellation of constant pre-factors:
\ba
\label{sol2}
\beta &=& 2 \alpha =\frac1{c_s^2}+1\, , \\
v &=&  - \dfrac{M^4}{c_s^2} \left( \dfrac{V_0 \kappa^2}{6 M^4} \right)^\alpha \left(1+\dfrac{3 c_s^2}{\kappa} \right) \, .
\ea
In addition, we also have
\ba
\epsilon = \dfrac{XP_{,X}}{\mPl^2 H^2} \propto a^{2 \alpha \kappa} \, .
\ea
Recalling $\epsilon\propto a^\eta$, the  parameter $\kappa$ being a
constant is consistent with the parameter $\eta$ being a constant,  as
desired.  We find
\ba
\label{sol3}
\kappa \simeq \dfrac{\eta}{2 \alpha} \, .
\ea
We have five free parameters in our action. In addition, $\kappa$
is another parameter obtained from the
solution. Among all, two of them are determined by requiring the ansatz
given in \eqref{ansatz} to be a consistent solution, and two others are
fixed for
a given value of $\eta$ and $c_s$. In the end, two parameters remain
undetermined. As we shall show in Section~\ref{linear-perturbation}, a
scale-invariant
power spectrum requires $\eta \simeq -6$, and a large non-Gaussianity
requires  $c_s\ll 1$.

Using Eq. \eqref{sol1} and Eq. \eqref{sol2}, one can easily check that
\ba
\dfrac{X^\alpha/M^{4 \alpha-4}}{v \left(\phi/\mPl \right)^\beta} \simeq - c_s^2 \left( 1+\dfrac{3c_s^2}{\kappa} \right)^{-1} \, .
\ea
As a result, for $c_s \ll 1$, the kinetic term is always sub-dominant in comparison to the potential term.

In the above calculations, we have assumed that the term linear in $X$
is negligible, i.e., $X \ll X^\alpha/M^{4\alpha -4}$, or equivalently
$(X/M^4)^{\alpha-1}\gg 1$. For $\alpha\gg 1$ (i.e., $c_s\ll 1$), this
implies $X/M^4>1$. Using the ansatz given in \eqref{ansatz}, this condition
translates to
\ba
\sqrt{\dfrac{V_0}{6\mPl^2}} \dfrac{\vert \kappa \vert \phi}{ M^2} > 1 \, ,
\ea
and the condition breaks down at $\phi= \phi_*$ or $t=t_*$ defined by
\ba
\label{phi-star}
\frac{\phi_*}{\mPl} \simeq \sqrt{\dfrac{6}{V_0 }}\dfrac{M^2 }{\vert
\kappa \vert} \, .
\ea
After $\phi_*$, we enter the slow-roll inflation phase for a
relatively large range of initial conditions. If this does not happen,
we loose our analytic control on the solution and the curvature
perturbation may not be conserved in the second phase. Therefore, in what
follows, we shall assume that the second inflationary phase does occur
and is in the slow-roll regime. Furthermore, as $v<0$ for
$\eta \simeq -6$, we need
another phase to have a graceful exit from inflation before the negative
potential dominates. This can be achieved by coupling $\phi$
to another heavy (waterfall) field, as in hybrid inflation models
\cite{Linde:1993cn}.
Note that the waterfall field is needed only for
ending inflation and does not contribute to
super-horizon fluctuations, and thus our model remains a single-field model.

In this set up of the model,  the field climbs up the potential during
the first phase
of inflation. This is why we have a non-attractor background initially.
Note that the ansatz given in Eq. \eqref{ansatz} and the fine-tuning between
parameters given in Eq. \eqref{sol2} and Eq. \eqref{sol3} are {\it not}
necessary conditions for obtaining the  non-attractor behavior. We require these
specific values of the parameters in order to be able to do analytic
calculations of the bispectrum. In fact, we find a non-attractor phase
for a half of the ranges of possible initial conditions.

\begin{figure}
\includegraphics[ scale=.5]{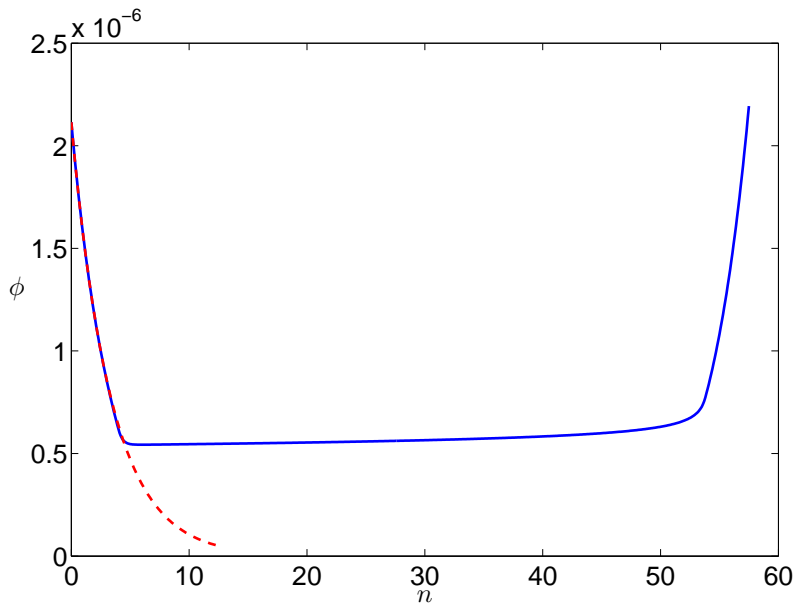}
\caption{Evolution of $\phi (n)$ in the undershoot situation as a
 function of the number of $e$-folds, $n$, counted from the beginning of
 inflation. The inflaton field climbs up the potential, stops somewhere
 before the top of the potential ($\phi=0$), turns around and goes back
 to plus infinity.
The dashed red curve is the analytic ansatz for the non-attractor phase,
 while the solid blue curve is the full numerical solution.
The transition to a slow-roll inflation phase is sharp, and
 an extended  slow-roll phase follows afterward. The parameters,
 consistent with Eqs.~\eqref{sol2} and \eqref{sol3}, in units of $M_P$ are
 $V_0 \simeq 6.25 \times 10^{-4}$, $M=5 \times 10^{-5}$, $\alpha = 10$,
 and $\eta = -6$.}
\label{stop}
\end{figure}

\begin{figure}
\includegraphics[ scale=.5 ]{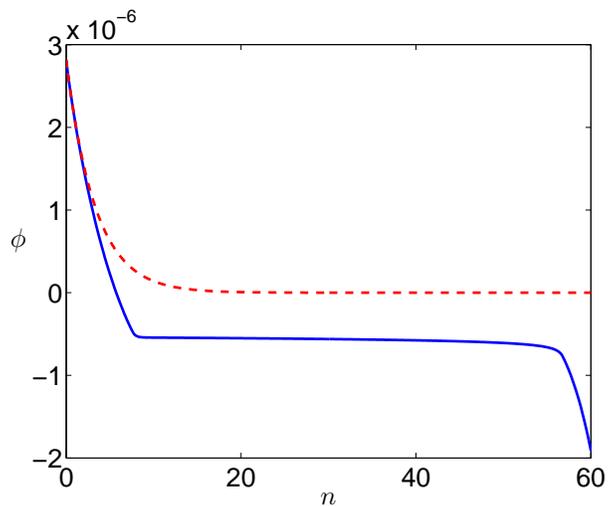}
\caption{Same as Fig. \ref{stop}, but for the overshoot
 situation. The inflaton field climbs up the potential, goes over the
 top of the potential, and rolls down on the other side of the
 potential.
}
\label{cross}
\end{figure}

Depending on initial conditions, the solution for $\phi$ shows three
different behaviors: the undershoot, the critical or the overshoot. In the
undershoot case, the inflaton field
climbs up the potential, stops somewhere before crossing the origin (the
top of the potential), turns around and rolls down on the same side of
the potential. In this case $\phi$ always has a unique sign while $\dot
\phi$ changes the sign (see Fig. \ref{stop}).
In the overshoot case, the inflaton
field climbs up the potential with a large enough initial velocity, so
that it goes
over the top of the potential, and rolls down on the other side of the
potential.  In this case $\phi$ changes the sign while $\dot \phi$
always has a unique sign (see Fig. \ref{cross}).
The critical limit occurs when the initial conditions are such that it takes
infinite amount of time for the inflaton field to reach the top of the
potential.

%%%%%%%%%%%%%%%%%%%%%%%%%%%%%%%%%%%%%%%%%%%%%%%%%%
%%%%%%%%%%%%%%%%%%%%%%%%%%%%%%%%%%%%%%%%%%%%%%%%%%

%%%%%%%%%%%%%%%%%%%%%%%%%%%%%%%%%%%%%%%%%%%%%%%%%%
%%%%%%%%%%%%%%%%%%%%%%%%%%%%%%%%%%%%%%%%%%%%%%%%%%

\begin{figure}
\includegraphics[ scale=.55 ]{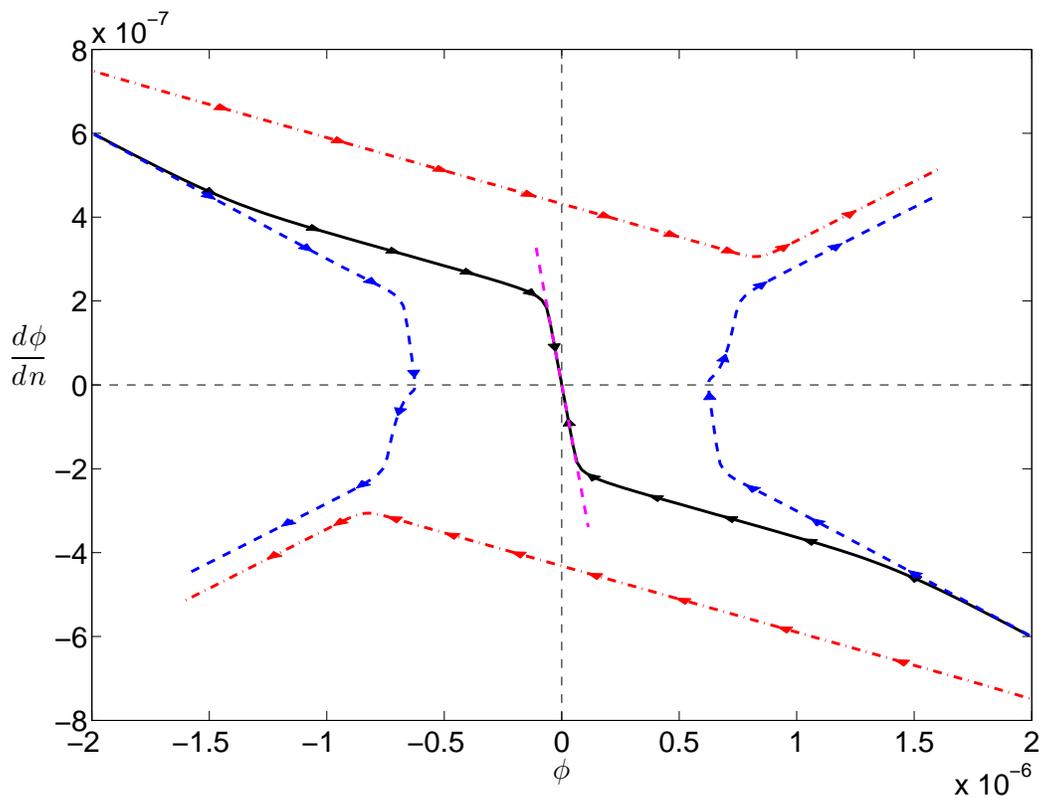}
\caption{Phase-space diagram of the model given by
 Eq.~\eqref{model}. The black line separates two different trajectories:
 the undershoot (dashed blue lines) and overshoot (dot-dashed red
 lines) trajectories. The purple dashed line near the origin shows the
 asymptotic
 solution, $\phi \propto a^{-3}$. In the undershoot case, the inflaton
 field climbs up the potential, stops somewhere before reaching the top
 of the potential, and returns back. In the overshoot case, the field
climbs up the potential,  crosses the top of the  potential, and
rolls down on the other side of the potential. The symmetry in the plot
 reflects the fact that our Lagrangian given in Eq.~\eqref{model} is
 symmetric under
 the transformation $\phi
\to - \phi $ and $\dot \phi \to - \dot \phi $.	 }
\label{phase}
\end{figure}

These different behaviors of the inflaton-field evolution in phase space
are shown in Fig. \ref{phase}. The critical limit (the black solid line
in  Fig. \ref{phase}) separates the overshoot and undershoot solutions.
The early-time behavior of this curve for large $|\phi|$ and $|\dot
\phi|$ (for which the power-law term, $X^\alpha$, dominates) is
asymptotically the same as the ansatz we obtained above. On the other
hand, the linear
term in $X$ dominates near the origin, and one should solve the
equation of motion for a
canonically normalized field, i.e.,
\ba
\label{linear}
\ddot \phi + 3 H \dot \phi + \dfrac{v\beta}{\mPl} \left(\dfrac{\phi}{\mPl} \right)^{\beta -1} = 0\, .
\ea
For $\beta\gg 1$, the last term proportional to
$(\phi/\mPl)^{\beta-1}$ is small relative to the first two terms near
the origin, $\phi/\mPl\simeq  0$. Thus,
the slow-roll condition no longer holds, and
we have $\ddot \phi + 3 H \dot \phi \simeq 0$. This is similar to the
scenario of a constant potential studied in Ref.~\cite{Namjoo:2012aa}.  The
solution is  $\phi \propto a^{-3}$ and, as a result, $d \phi / dn \simeq
-3 \phi$, where $n$ is the number of $e$-folds counted from the beginning
of inflation. This asymptotic solution  is in agreement with the
numerical one for $\phi/\mPl\simeq  0$ (see the purple line in
Fig.~\ref{phase}).

%%%%%%%%%%%%%%%%%%%%%%%%%%%%%%%%%%%%%%%%%%%%%%%%%%
%%%%%%%%%%%%%%%%%%%%%%%%%%%%%%%%%%%%%%%%%%%%%%%%%%

%%%%%%%%%%%%%%%%%%%%%%%%%%%%%%%%%%%%%%%%%%%%%%%%%%
%\section{The linear perturbation}
\section{Power spectrum}
\label{linear-perturbation}

In this section, we calculate the power spectrum of curvature
perturbations generated during the non-attractor phase and obtain the
condition for a scale-invariant power spectrum.
As usual, we have the following quadratic action for curvature perturbation \cite{Mukhanov:1985rz,Sasaki:1986hm,Garriga:1999vw}:
\label{pert}
\ba
S=\dfrac{1}{2} \int d^3 x d \tau~ z^2 \left[ \calR'^2 - c_s^2 (\nabla
\calR)^2 \right]\, ,
\ea
where
\ba
z^2 \equiv \dfrac{2\epsilon a^2}{c_s^2} \mPl^2\, .
\ea
Recalling $\epsilon \propto a^\eta$ and assuming a
Bunch-Davies initial state deep inside the horizon, the solution for the
mode function is given by
\ba
 {\cal R}_k = C x^\nu H_\nu^{(1)}(x)\, ,
\ea
where we have defined
\ba
x\equiv- c_s k \tau  \qquad , \qquad \nu\equiv \dfrac{3+\eta}{2}\, ,
\ea
$\tau$ is the conformal time defined by $d\tau \equiv dt/a(t)$, and
\ba
\label{C}
\vert C \vert^2 \equiv \dfrac{\pi c_s}{8 k \epsilon_i a_i^2 \mPl^2}
x_i^{1-2\nu} \, .
\ea
The subscripts $i$ denote the corresponding values at the start of inflation.

The power spectrum of curvature perturbations at the end of
the non-attractor phase is given by
\ba
{\cal P_R}= \dfrac{k^3}{2 \pi^2} \vert \calR_k \vert^2\, .
\ea
Note that this power spectrum will be the observed power
spectrum, as ${\cal R}$ is conserved outside the horizon  after the end
of the non-attractor phase.

Using Eq.~\eqref{C} we write the power spectrum in terms of the
parameters at the end of the non-attractor phase as
\ba
{\cal P_R} \simeq \dfrac{\Gamma(\vert \nu \vert)^2 }{\pi^3 2^{2\nu+4} }
\left(\dfrac{H_*}{\mPl} \right)^2 \dfrac{1}{c_s\epsilon_*} \left(
\dfrac{c_sk}{H_* a_*} \right)^{3+2\nu}\, ,
\label{PR-star}
\ea
where we have assumed $\nu=(3+\eta)/2<0$, so that we can expand the
Hankel function for a small argument, $x\ll 1$.

During the non-attractor phase, the fast decay of $\epsilon$ makes the
curvature perturbation grow very rapidly on super-horizon
scales. This growth continues until the end of the
non-attractor phase. The subsequent slow-roll phase begins at $t=t_*$ or
$\phi=\phi_*$ given in Eq. (\ref{phi-star}). The curvature perturbation
is conserved during the slow-roll phase.

The spectral index is given by
\ba
n_s-1 \simeq 3+2\nu = 6+\eta\, .
\ea
Therefore,  $\eta = -6$ is required for a scale-invariant power
spectrum. A slightly red-tilted  power spectrum, $n_s=0.96$
\cite{Hinshaw:2012fq,Ade:2013uln}, can be
easily obtained by choosing $\eta= -6.04$.

%%%%%%%%%%%%%%%%%%%%%%%%%%%%%%%%%%%%%%%%%%%%%%%%%%
\section{Bispectrum: in-in formalism}
\label{in-in}
In this section, we calculate the bispectrum of curvature
perturbations generated during the non-attractor phase using  the in-in
formalism in two different gauges. We provide the detailed
derivations of the results presented earlier in Ref.~\cite{ Chen:2013aj}.

 The first gauge is the comoving gauge, which enables the most
 complicated but rigorous calculations.
The bispectrum we obtain in the comoving gauge is valid
 for arbitrary values of $c_s$, including $c_s=1$.
The second gauge is the flat
 gauge, and we use this gauge to show that a large bispectrum
 comes from the matter sector. As we use the
 decoupling-limit approximation when computing the bispectrum in the
 flat gauge, the bispectrum
 in this gauge is valid only for $c_s\ll 1$.
 We also
 explicitly show that the results from two gauges are equivalent to each
 other in the small sound-speed limit.

%%%%%%%%%%%%%%%%%%%%%%%%%%%%%%%%%%%%%%%%%%%%%%%%%%
\subsection{Comoving gauge}
\label{Sec:comoving}

The cubic action in the comoving gauge, in which the scalar
field is unperturbed, is given by \cite{Chen:2006nt,Seery:2005wm}
\begin{eqnarray} \label{action3}
S_3&=&\int dt d^3x\left\{
-a^3 \left[ \Sigma\left(1-\frac{1}{c_s^2}\right)+2\lambda
     \right]\frac{\dot{\calR}^3}{H^3}
+\frac{a^3\epsilon}{c_s^4}(\epsilon-3+3c_s^2)\calR\dot{\calR}^2 \right.
\nonumber \\ &+&\left.
\frac{a\epsilon}{c_s^2}(\epsilon-2s+1-c_s^2)\calR(\partial\calR)^2 \right.
\nonumber \\
&-& 2a \frac{\epsilon}{c_s^2}\dot{\calR}(\partial
\calR)(\partial \chi)
\nonumber \\
&+& 2 f(\calR)\left.\frac{\delta L}{\delta
\calR}\right|_1
\nonumber \\
&+&\left.
\frac{a^3\epsilon}{2c_s^2}\frac{d}{dt}\left(\frac{\eta}{c_s^2}\right)\calR^2\dot{\calR}
+\frac{\epsilon}{2a}(\partial\calR)(\partial
\chi) \partial^2 \chi +\frac{\epsilon}{4a}(\partial^2\calR)(\partial
\chi)^2\right\} ~,
\end{eqnarray}
where
\begin{eqnarray}
\label{chidef}
\partial^2\chi \equiv a^2\epsilon\dot{\cal R}~,
\end{eqnarray}
\begin{eqnarray}
\left.\frac{\delta
L}{\delta\calR}\right|_1 \equiv a
\left( \frac{d\partial^2\chi}{dt}+H\partial^2\chi
-\epsilon\partial^2\calR \right) ~,
\end{eqnarray}
\begin{eqnarray} \label{redefinition}
f(\calR)&\equiv& \frac{\eta}{4c_s^2}\calR^2+\frac{1}{c_s^2H}\calR\dot{\calR}+
\frac{1}{4a^2H^2}[-(\partial\calR)(\partial\calR)+\partial^{-2}(\partial_i\partial_j(\partial_i\calR\partial_j\calR))] \nonumber \\
&+&
\frac{1}{2a^2H}[(\partial\calR)(\partial\chi)-\partial^{-2}(\partial_i\partial_j(\partial_i\calR\partial_j\chi))] ~.
\end{eqnarray}
The terms in the last line in Eq.~\eqref{action3} are
higher-order in
$\epsilon$ and can be ignored. In the usual attractor inflation models,
for which $\dot{\cal R}\approx 0$ on super-horizon scales and thus the
contributions to the integral come from the horizon-crossing epoch,
$kc_s\approx aH$, all of the terms in the first two lines in
Eq.~\eqref{action3}, which are proportional to $\dot{\cal R}^3$, ${\cal
R}\dot{\cal
R}^2$, ${\cal R}(\partial {\cal R})^2$, yield the equilateral bispectrum
\cite{Seery:2005wm,Chen:2006nt}.
However, in this non-attractor model, for which $\dot{\cal R}=3H{\cal
R}$ on super-horizon scales, the integral receives dominant
contributions after the horizon exit. As a result, the terms in the
first line, $\dot{\cal R}^3$ and ${\cal R}\dot{\cal R}^2$, are
proportional to  ${\cal R}^3$ on super-horizon scales,  yielding
the local-form bispectrum.
%This observation also suggests that the terms
%in the second line in Eq.~\eqref{action3}, ${\cal R}(\partial {\cal
%R})^2$, are
%sub-dominant, as they do not contribute on super-horizon scales unlike
%the terms in the first line.
Also in this non-attractor case, all the terms with spatial derivatives are suppressed by factors of the scale factor at the end of the non-attractor inflationary phase, $k/(a_{\rm end} H)$, and negligible.

Therefore, we have a rather different
situation here: in the usual attractor case, all the terms in the first
three lines in Eq.~\eqref{action3} must be included for consistent
computation of the bispectrum up to $f_{NL} \sim {\cal O}(\epsilon)$, whereas in the non-attractor case only the terms in the first line are
necessary. Note that this statement is independent of the value of
$c_s$, and thus the results given in this section are valid for arbitrary
values of $c_s$, including $c_s=1$.

In the usual attractor case, the terms in the first two lines in
Eq.~\eqref{action3} give the equilateral bispectrum with $f_{NL}^{\rm
equil}={\cal O}(1/c_s^2)$ for $c_s\ll 1$
\cite{Chen:2006nt}. In the non-attractor case, the terms in
the first line give a large local bispectrum with $f_{NL}^{\rm
local}={\cal O}(1/c_s^2)$ for $c_s\ll 1$ as we shall show below, whereas
the other terms are negligible.
Therefore, the non-attractor model yields un-observable signals in the
equilateral bispectrum.

How about the fourth line in Eq.~\eqref{action3}, which can be removed by
a field redefinition? Again, we only need to keep
the terms that do not have extra spatial derivatives. Ignoring the terms
suppressed by spatial derivatives in Eq.~\eqref{redefinition}, we redefine the curvature perturbation as
\ba
\calR \to \calR_n + \dfrac{\eta}{4 c_s^2} \calR_n^2 + \dfrac{1}{c_s^2 H}
\calR_n \dot \calR_n \, ,
\label{field_redef}
\ea
where ${\cal R}_n$ is the redefined field.
After the horizon crossing, when the argument of the Hankel function with
rank $\nu<0$ is small, we have
\ba
\calR' =-c_s k x^\nu H^{(1)}_{\nu-1}(x)\simeq -c_s k \dfrac{2 \nu}{x} \calR\, .
\ea
As a result, the above field redefinition becomes
\ba
 \calR \to \calR_n + \left(\dfrac{\eta}{4 c_s^2}-\dfrac{2 \nu}{c_s^2}
 \right)\calR_n^2 \, .
\label{field_redef2}
\ea
The quadratic terms in Eq.~\eqref{field_redef2} give the
following contribution to the local-form $f_{\rm NL}$ parameter (denoted as
``$f_{NL}^{\mathrm{FR}}$''):
\ba
\label{fNLFR}
\dfrac{3}{5} f_{NL}^{FR} = \dfrac{1}{4 c_s^2} (\eta - 8 \nu)
= - \dfrac{1}{4 c_s^2} (12+3\eta) \, .
\ea

The first term in Eq.~\eqref{action3} gives the bispectrum of
\ba
\langle \calR_{\bf k_1} \calR_{\bf k_2} \calR_{\bf k_3} \rangle_{\dot \calR^3} = 6 \times 2 {\rm M^2_P} \,
{\rm Im} \left[ \calR_{k_1}(\tau_*) \calR_{k_2}(\tau_*) \calR_{k_3}(\tau_*)
\int_{-\infty}^{\tau_*} d\tau \, [\Sigma (1-1/c_s^2) + 2 \lambda] \calR_{k_1}'^*(\tau) \calR_{k_2}'^*(\tau) \calR_{k_3}'^*(\tau)
\right] \, ,
\ea
where $\tau_*$ is the conformal time at the end of the non-attractor
phase. As the kinetic term is dominated by $X^\alpha$ during the
non-attractor phase, $\lambda$ is given by
\ba
\lambda = \dfrac{\Sigma}{6} \left(\dfrac{1}{c_s^2}-1\right) \, ,
\ea
where we have used Eq.~\eqref{sound-speed}. Recall that
$\Sigma=H^2\mPl^2\epsilon/c_s^2$.
Ignoring a small tilt and setting  $\eta = -6$ and $\nu = -3/2$, the
mode function simplifies to
\ba
\calR_k = C_k \sqrt{\dfrac{2}{\pi}} \, \dfrac{e^{-i c_s k \tau}}{(-c_s k \tau)^3} (-1-i c_s k \tau) \, .
\label{mode_function}
\ea
As a result, the first term in Eq.~\eqref{action3} gives the local-form
bispectrum parameter of
\ba
\dfrac{3}{5} f_{NL}^{\dot \calR^3} = -\dfrac{3}{2 c_s^2} (1-c_s^2) \, .
\ea
With a similar procedure, the second term gives the local-form
bispectrum parameter of
\ba
 \dfrac{3}{5} f_{NL}^{\calR \dot \calR^2} = \dfrac{3}{4 c_s^2} (1-c_s^2)\, .
\ea

The total local-form bispectrum parameter, $f_{NL}^{\mathrm{local}}$, is
 given by the sum of the above contributions:
\ba
\label{fNL-comoving}
  \dfrac{3}{5} f_{NL}^{\mathrm{local}} =  \frac{3}{5} \left( f_{NL}^{\dot \calR^3} + f_{NL}^{\calR \dot \calR^2} + f_{NL}^{FR} \right) =
  \dfrac{3}{4 c_s^2} (1+c_s^2) \, .
\ea
Once again, this result is valid for arbitrary values of
 $c_s$, including $c_s=1$.
As emphasized in Ref.~\cite{ Chen:2013aj}, Eq.~\eqref{fNL-comoving}
shows that  the presence of a large primordial $f_{NL}^{\rm local}$ would
not rule out all single-field models in full
generality. Rather, it would rule out
all single-field models which have reached the attractor solution and with
a Bunch-Davies initial state.\footnote{Also see the workshop summary
of ``Critical Tests of Inflation Using Non-Gaussianity'' in \url{http://www.mpa-garching.mpg.de/~komatsu/meeting/ng2012/}.}

As we have shown above, this model gives a local-form
bispectrum because $\dot{\cal R}^3$ and ${\cal R}{\dot{\cal R}^2}$
become proportional to ${\cal R}^3$ on super-horizon scales. This
implies that we can obtain the same result using classical calculations
such as the $\delta N$ formalism, which uses gradient expansion. We
shall confirm this in Section~\ref{delta-N}.

The subsequent slow-roll phase of inflation after the first
non-attractor phase cannot change the value of $f_{NL}^{\rm local}$, as the
super-horizon curvature perturbation remains constant during
the slow-roll phase.
However, one may wonder what would happen to $f_{NL}^{FR}$, i.e., the
contribution from the field-redefinition terms given in
Eq.~\eqref{field_redef}, which are suppressed during the
slow-roll phase
by $\eta\ll 1$ and $\dot{\calR} \approx 0$.
While it is true that the field-redefinition terms become
negligible during the slow-roll phase, we find that, in the comoving
gauge, a boundary term in the cubic action at the end of the
non-attractor phase replaces the contributions from the
field-redefinition terms.

To show this explicitly, let us model the evolution of
$\eta$ such that it is equal to $\eta=\eta_0=-6$ during the non-attractor phase
and vanishes during the slow-roll phase. Specifically,
\ba
\eta = \eta_0 \left( 1-\theta(t-t_*) \, \right)\, ,
\ea	
where $t_*$ is the transition time at which $\phi(t_*)=\phi_*$ given in Eq. (\ref{phi-star}).
This step function becomes a delta function upon  a time derivative with
respect to $t$. As a result, this gives a boundary term
in the cubic action which has a non-negligible contribution:
\ba
S_3\ni
\int d\tau d^3x \, \frac{a^2\epsilon}{2 c_s^2} \dfrac{d}{d\tau} \left( \dfrac{\eta}{c_s^2} \right) \calR^2 \calR'	
 \simeq \int d^3x\left[ \dfrac{a^2 \epsilon}{2 c_s^4} \eta_0 \calR^2 \calR' \right]_{*}\,.
\ea
One can check that $f_{NL}^{\rm local}$ from this term is
equal to $f_{NL}^{FR}$ given by Eq.~\eqref{fNLFR}. Therefore, this term
replaces $f_{NL}^{FR}$ after the end of the non-attractor phase, and the
total $f_{NL}^{\rm local}$ remains equal to that given by
Eq.~\eqref{fNL-comoving}.

%%%%%%%%%%%%%%%%%%%%%%%%%%%%%%%%%%%%%%%%%%%%%%%%%%
\subsection{Flat gauge}
\label{Sec:flat}
In the small sound-speed limit, the bispectrum is sourced
primarily by interactions in the scalar-field sector, and the
interactions involving gravity become negligible. In such cases, it
is known that the computation of the bispectrum
can be made simpler by using the so-called ``inflaton
approximation'' or the ``decoupling limit''
\cite{Creminelli:2003iq,Alishahiha:2004eh,Gruzinov:2004jx,Chen:2009bc,Chen:2009zp,Leblond:2010yq}.
In this approximation, we ignore metric perturbations
entirely, and consider only the scalar-field perturbation:
\bea
\phi(\bx,t) = \phi_0(t) + \delta\phi(\bx,t) ~.
\eea
To derive the cubic action in $\delta\phi$, we simply perturb
$P(X,\phi)$ with respect to $X$, and obtain
\bea
\CL_3 = a^3 \frac{2\lambda}{\dot\phi_0^3} \delta{\dot\phi}^3
- a \frac{\Sigma (1-c_s^2)}{\dot\phi_0^3} \delta{\dot\phi} (\partial \delta\phi)^2 ~.
\label{action_deltaphi}
\eea
The perturbations of $P(X, \phi)$ with respect to $\phi$ can be ignored,
as they are not enhanced by $c_s^{-2}$ or $\lambda/\Sigma$.
While we shall loosely call this action the ``cubic action
in the flat gauge'' in this paper, this action is not the full cubic
action in the flat gauge, as we have ignored terms coming from the lapse
function and the shift vector via the constraint equations (Lagrange
multipliers). Once again, ignoring these terms and working only with the
above two terms is justified only in the decoupling limit, in which the
scalar-field interactions overwhelm the gravitational ones.

Using the relation $\calR =-H\delta\phi/\dot\phi$, we rewrite this action as
\bea
\CL_3 = -2a^3 \frac{\lambda}{H^3} \dot\calR^3
+ a \frac{\Sigma (1-c_s^2)}{H^3} \dot\calR (\partial \calR)^2 ~.
\label{action_deltaphi2}
\eea
In Appendix \ref{App:Equiv}, we show that the action given in
Eq.~(\ref{action_deltaphi2}) is
equivalent to that in the comoving gauge in the leading order
of $\lambda/\Sigma$ and $c_s^{-2}$, and for $H$,
$\dot\phi$, $\eta$, $c_s \sim {\rm const.}$, including the
field-redefinition terms.

In our model, the field interactions build up on super-horizon
scales. The second term in Eq.~\eqref{action_deltaphi} is thus
subdominant due to the spatial derivative, and we only need to compute the
first term. We obtain
\ba
\label{fNL-flat}
\frac{3}{5} f_{NL}^{\rm local} \simeq \dfrac{3}{4} \left(\frac{1}{c_s^2}
-1\right) ~.
\ea
As expected, for $c_s\ll 1$, this simple method reproduces the leading order result of the previous section, i.e., Eq. (\ref{fNL-comoving}).

%The computation so far gives us the following picture for the origin of
%this large $f_{NL}^{\rm local}$. The mode function given by
%Eq.~\eqref{mode_function} is growing on super-horizon scales, and the
%interactions of the scalar field on super-horizon scales are responsible
%for this  large $f_{NL}^{\rm local}$. These properties suggest that
%we should be able to obtain the same results using the classical
%calculations without using quantum field theory calculations such as the
%in-in formalism. This is indeed what we shall find in Section~\ref{delta-N}
%using the $\delta N$ formalism.

%%%%%%%%%%%%%%%%%%%%%%%%%%%%%%%%%%%%%%%%%%%%%%%%%%
\section{$\delta N $ formalism}
\label{delta-N}

In this section we shall use the $\delta N$ formalism
\cite{Starobinsky:1982ee,Sasaki:1995aw,Sasaki:1998ug,Wands:2000dp,
Lyth:2004gb, Lyth:2005fi}
to calculate $f_{NL}^{\rm local}$.  We shall show that $f_{NL}^{\rm local}$ we
have calculated using the in-in formalism in Section~\ref{in-in} agrees
precisely with that we find from the $\delta N$ formalism in this
section. As we have shown already in Section~\ref{in-in},
this is because the intrinsic bispectrum of the quantum
fluctuations present at the time of the horizon crossing is
sub-dominant, and the dominant contribution comes from the interactions
of the scalar field on super-horizon scales. Fortunately this is all one
needs in  using the $\delta N$ formalism based on a separate universe
assumption \cite{Lyth:2004gb} (see \cite{Sugiyama:2012tj} for more
precise conditions under which the $\delta N$ formalism is valid).

Nevertheless, extra cares must be taken when we use the $\delta N$
formalism in non-attractor backgrounds. Once the solution
reaches the attractor solution, we need to consider only the
perturbations of the scalar-field
trajectories with respect to the field value at the initial
hypersurface, $\phi$, as the velocity, $\dot\phi$, is uniquely
determined by $\phi$. However, in the non-attractor case, 
the scalar-field trajectories are not uniquely
determined by the field value $\phi$ alone. We also need the
information of $\dot \phi$ to determine the trajectory~\cite{Namjoo:2012aa}.

In order to find the scalar-field trajectories, we need to solve the equation
of motion of the scalar field, which is a second-order differential equation. 
We thus need to provide two initial conditions ($\phi$ and $\dot\phi$) on the
initial hypersurface. We can then integrate the equation of motion to
the final time, $t=t_*$. We assume that the universe has already arrived 
at the attractor phase (often called the adiabatic limit) by this epoch,
or a phase transition to an attractor phase occurs at $t=t_*$. 
More specifically, we assume that the evolution of the universe is unique
after the value of the scalar field has arrived at $\phi=\phi_*$, 
irrespective of the value of its velocity $\dot\phi_*$. In other words, 
at and after $t=t_*$, the scalar field plays the role of a clock. 
We note that this is a necessary condition for the validity of the 
$\delta N$ formalism, since only in this case $\delta N$ is equal to
the final value of the comoving curvature perturbation $\calR$ which
is conserved at $t\geq t_*$.
Thus the number of $e$-folds $N$ counted backward from
the epoch when $\phi=\phi_*$ to an earlier epoch is a function of
$\phi$ and $\dot\phi$, $N=N(\phi,\dot\phi;\,\phi_*)$.

With this in mind we apply the $\delta N$ formalism.
Our program is as follows. In order to find the background
scalar-field trajectories, we solve the equation of motion of the scalar
field perturbatively by expanding it around a particular trajectory given by
$\phi\propto e^{\kappa Ht}$. We then use these background solutions for
the field trajectories to compute the perturbations of the number of
$e$-folds with respect to the initial field value and its time derivative.

%%%%%%%%%%%%%%%%%%%%%%%%%%%%%%%%%%%%%%%%%%%%%%%%%%
\subsection{The case with $c_s=1$}\label{cs1}

To familiarize ourselves with the $\delta N$ calculation in
 the non-attractor background, let us first work out the simplest case
 with the canonical
kinetic term, $c_s=1$. During the non-attractor phase whose potential is
dominated by a constant term, the background Klein-Gordon equation is
given by
\ba
\ddot \phi + 3H\dot \phi = 0 \, ,
\ea
which has the following solution
\ba
\phi = \lambda + \mu e^{-3Ht}\, ,
\ea
where $\lambda$ and $\mu$ are constants of 
integration.\footnote{We change the notation. Henceforth,
$\lambda$ is not given by Eq.~\eqref{lambda}, but is a constant of
integration.}
Without loss of generality, we assume $\dot\phi>0$.
We set $\phi(t_*)=\phi_*$ at which the non-attractor phase ends.

The number of $e$-folds counted backward in time from $t=t_*$ is
\be
N=\int_{t}^{t_*}Hdt=H(t_*-t)=-Ht\,,
\ee
where we have set $t_*=0$ without loss of generality.
With this definition of time, the above solution becomes
 \ba
\phi = \lambda + \mu e^{3N} = \lambda + (\phi_*-\lambda)e^{3N}\, .
\label{solphi}
\ea
This gives
\begin{eqnarray}
\dot\phi=-3H\mu e^{3N}=3H(\lambda-\phi_*)e^{3N}\,.
\label{soldotphi}
\end{eqnarray}

As clear from the above, the different trajectories in the phase 
space $(\phi,\dot\phi)$ are parameterized by $\lambda$ with $N$ being
the parameter along each trajectory. That is,
\begin{eqnarray}
\phi=\phi(N,\lambda)\,,
\quad
\dot\phi=\dot\phi(N,\lambda)\,.
\end{eqnarray}
In other words, the variables $(N,\lambda)$ may be regarded as 
another set of coordinates in the phase space. Thus one can
invert the above to obtain $N$ and $\lambda$ as functions of $(\phi,\dot\phi)$.
Specifically we obtain
\begin{eqnarray}
N&=&N(\phi,\dot\phi)
=\frac{1}{3}\ln\left(\frac{\dot\phi}{\dot\phi+3H(\phi-\phi_*)}\right)\,,
\label{Nsol}
\\
\lambda&=&\lambda(\phi,\dot\phi)=\phi+\frac{\dot\phi}{3H}\,.
\label{lambdasol}
\end{eqnarray}

In the present case of $c_s=1$, Eq.~(\ref{Nsol}) for $N$ in terms
of $\phi$ and $\dot\phi$ is sufficient to derive the $\delta N$ formula.
Nevertheless, for the sake of the discussion in the next subsection in which 
the case with $c_s\neq1$ is considered, we insert an intermediate step for the
derivation of the $\delta N$ formula as follows.

In place of $(\phi,\dot\phi)$, we may introduce a yet another set of
coordinates in the phase space. Here we choose $(\phi,\lambda)$.
A special feature of this choice is that one of the coordinates $\lambda$
is a constant of integration along each trajectory. Therefore, in particular,
its perturbation $\delta\lambda$ can be evaluated at any point along the
trajectory. With this choice, we have $N=N(\phi,\lambda)$. This expression
can be immediately obtained by inverting the solution of $\phi$ given by
Eq.~(\ref{solphi}), 
\ba
\label{simplesolution}
N=\frac13\ln\left(\frac{\phi-\lambda}{\phi_*-\lambda}\right)\, .
\ea
Then one may expand this by setting $\phi\to\phi+\delta\phi$
and $\lambda\to\lambda+\delta\lambda$, 
Up to the second order, we have
\begin{eqnarray}
\delta N &=&
\frac{\partial N}{\partial \phi}\delta\phi
+ \frac{\partial N}{\partial \lambda}\delta\lambda
+ \frac12\frac{\partial^2 N}{\partial \phi^2}\delta\phi^2
+ \frac12\frac{\partial^2 N}{\partial \lambda^2}\delta\lambda^2
+ \frac{\partial^2 N}{\partial \phi\partial\lambda}\delta\phi\delta\lambda\,.
\end{eqnarray}

Now we identify the perturbations $\delta\phi$ and $\delta\lambda$ with
those evaluated on the flat hypersurface at or after which the scale of interest
has crossed out of horizon. For $\delta\phi$, this is immediate. As for
$\delta\lambda$, however, we need its relation to $\delta\phi$ and $\delta\dot\phi$.
In the present case, we can readily find this from Eq.~(\ref{lambdasol}),
\begin{eqnarray}
\delta\lambda=\delta\phi+\frac{\delta\dot\phi}{3H}\,.
\end{eqnarray}
If we recall that the quantum fluctuations are dominated by the
constant mode $\delta\phi= const.$ on superhorizon scales,
we immediately obtain $\delta \dot \phi=0$, and hence $\delta\lambda=\delta\phi$.
Inserting this to Eq.~(\ref{simplesolution}), we finally obtain
\begin{eqnarray}
\delta N&=&
\dfrac{\delta \phi}{3 (\phi_*-\lambda)}
+  \dfrac{\delta \phi^2}{6 (\phi_*-\lambda)^2}\,.
\end{eqnarray}
This $\delta N$ yields the following $f_{NL}^{\mathrm{local}}$:
\ba
 f_{NL}^{\mathrm{local}} =\dfrac{5}{2}\,.
\ea
This is of course in agreement with the result obtained 
by differentiating $N$ directly with respect to
$\phi$ and $\dot\phi$~\cite{Namjoo:2012aa}.

As mentioned in the above, for this particular setup, 
not only $\delta\lambda$ which is a constant of motion by definition
but also $\delta\phi$ is conserved on superhorizon scales, and 
$\delta\lambda=\delta\phi$. This implies that we may choose the
initial hypersurface to be infinitesimally close to the end of the 
non-attractor phase, i.e., $\phi\to \phi_*$. In other words,  
$\delta N$ is simply given by the difference in the number of $e$-folds between 
the flat and comoving slices at $t=t_*$, 
\begin{eqnarray}
\delta N &=&
\left.\frac{\partial N}{\partial \phi}\right|_* \delta\phi_*
+ \left.\frac{\partial N}{\partial \lambda}\right|_* \delta\lambda
+ \frac12\left.\frac{\partial^2 N}{\partial \phi^2}\right|_* \delta\phi_*^2
+ \frac12\left.\frac{\partial^2 N}{\partial \lambda^2}\right|_* \delta\lambda^2
+ \left.\frac{\partial^2 N}{\partial \phi\partial\lambda}\right|_*
\delta\phi_*\delta\lambda\,,
\end{eqnarray}
where $\delta\phi_*$ is the fluctuation evaluated on the flat slicing
at $t=t_*$.
We find that two of these derivatives, $\partial N/\partial\lambda|_*$ 
and ${\partial^2 N}/{\partial\lambda^2}|_*$, vanish, 
and thus we need to evaluate only the other three terms. The result is
\ba
\delta N=\dfrac{\delta \phi_*}{3 (\phi_*-\lambda)}
+  \dfrac{\delta \phi_*^2}{6 (\phi_*-\lambda)^2}\,.
\ea
Recalling $\delta\phi_*=\delta\phi=const.$, this 
again gives $f_{NL}^{\mathrm{local}} =5/2$. 

In general, provided that we know how $\delta\phi$ evolves on superhorizon scales,
we can obtain $\delta N$ by evaluating the fluctuations at $t=t_*$.
In this case, since we only need to know the dependence of the derivatives
of $N$ on $\phi$ and a constant of integration $\lambda$ at $t=t_*$,
the evaluation procedure can be simplified considerably.
We shall exploit this simplification in the next section where we deal 
with the case with $c_s\ne 1$.

%%%%%%%%%%%%%%%%%%%%%%%%%%%%%%%%%%%%%%%%%%%%%%%%%%
\subsection{The case with $c_s\ne 1$}

Having familiarized ourselves with the new $\delta N$
calculation for the
simplest case, let us now move onto the case with $c_s\ne 1$.
In what follows, we shall again neglect the canonical kinetic term
during the non-attractor phase for simplicity. The background equation
of motion is
\begin{eqnarray}
\label{KG}
\ddot\phi+3c_s^2H\dot\phi- F=0\, ,
\end{eqnarray}
where
\ba
F\equiv c_s^2\frac{P_{,\phi}}{P_{,X}} \, .
\ea
Finding a general analytical solution to this equation is
not easy. We thus first consider a particular solution,
$\phi=\phi_0\propto e^{\kappa Ht}$ (i.e., $\phi=\phi_0(N)=\phi_*e^{-\kappa N}$),
and then obtain a more general solution for the background  up to the
second order in perturbations around this particular solution. 
Here, as before, we assume that the non-attractor phase ends when $\phi=\phi_*$.
Using $\phi_0\propto e^{\kappa Ht}$ in $F$ yields
\begin{eqnarray}
F= c_s^2\frac{P_{,\phi}}{P_{,X}}
=F_0\left(\frac{\phi}{\phi_0}\right)^{2\alpha-1}
\left(\frac{\dot\phi}{\dot\phi_0}\right)^{2-2\alpha}\, ,
\end{eqnarray}
where
\begin{eqnarray}
F_0\equiv\ddot\phi_0+3c_s^2H\dot\phi_0=\kappa(\kappa+3c_s^2)H^2\phi_0\, .
\end{eqnarray}

Let us expand $F$ around $\phi=\phi_0$ to the second order. Defining
$$\chi\equiv\phi-\phi_0\, , $$ for notational
simplicity,\footnote{We change the notation. Henceforth,
$\chi$ is not given by in Eq.~\eqref{chidef}, but is the difference
between the true background solution and the reference solution,
$\chi\equiv \phi-\phi_0$.} the result is
\begin{eqnarray}
F&=&F_0\left[1+(2-2\alpha)\frac{\dot\chi}{\dot\phi_0}
+(2\alpha-1)\frac{\chi}{\phi_0}\right.
\cr
\cr
&+&\left.\frac{(2-2\alpha)(1-2\alpha)}{2}
\left(\frac{\dot\chi}{\dot\phi_0}\right)^2
+(2-2\alpha)(2\alpha-1)\frac{\dot\chi\,\chi}{\dot\phi_0\,\phi_0}
+\frac{(2\alpha-1)(2\alpha-2)}{2}
\left(\frac{\chi}{\phi_0}\right)^2
\right]
\cr
\cr
&=&
F_0\left[1+\frac{(2-2\alpha)}{\kappa}\frac{\dot\chi}{H\phi_0}
+(2\alpha-1)\frac{\chi}{\phi_0}\right.
\cr
\cr
&+&\left.\frac{(2-2\alpha)(1-2\alpha)}{2\kappa^2}
\left(\frac{\dot\chi}{H\phi_0}\right)^2
+\frac{(2-2\alpha)(2\alpha-1)}{\kappa}\frac{\dot\chi\,\chi}{H\phi_0^2}
+\frac{(2\alpha-1)(2\alpha-2)}{2}
\left(\frac{\chi}{\phi_0}\right)^2
\right]\,.
\label{F-expansion}
\end{eqnarray}
Having obtained $F$ in Eq. (\ref{KG}) to the linear and quadratic orders
in $\chi$, as given in Eq. (\ref{F-expansion}), we
are ready to solve Eq. (\ref{KG}) perturbatively.

%%%%%%%%%%%%%%%%%%%%%%%%%%%%%%%%%%%%%%%%%%%%%%%%%%
\subsubsection{Linear perturbation}

Let us consider the linear perturbation, $\chi_1$. The equation of
motion is
\begin{eqnarray}
0&=&\ddot\chi+3c_s^2H\dot\chi-F_0
\left[\frac{(2-2\alpha)}{\kappa}\frac{\dot\chi}{H\phi_0}
+(2\alpha-1)\frac{\chi}{\phi_0}\right]
\cr
\cr
&=&
\ddot\chi+\left[3c_s^2+(2\alpha-2)(\kappa+3c_s^2)\right]H\dot\chi
-(2\alpha-1)\kappa(\kappa+3c_s^2)H^2\chi
\cr
&=&
\ddot\chi+\left[3+(2\alpha-1)\kappa-\kappa\right]H\dot\chi
-\kappa\left[3+(2\alpha-1)\kappa\right]H^2\chi
\label{chi1-eq}\,.
\end{eqnarray}
The general solution is given by
\begin{eqnarray}
\chi=\chi_1\propto
\left\{\begin{array}{l}
\exp[\kappa Ht]\,,\\
\exp\left[-(3+\eta-\kappa) Ht\right]\,,
\end{array}
\right.
\label{linearsol}
\end{eqnarray}
where we have set $2\alpha\kappa=\eta$, and
$\eta=\dot\epsilon/H\epsilon$.
A scale-invariant spectrum requires $\eta \simeq -6$; thus, the second
solution, $\propto\exp[-(3+\eta-\kappa)Ht]$, will eventually dominate.

%%%%%%%%%%%%%%%%%%%%%%%%%%%%%%%%%%%%%%%%%%%%%%%%%%
\subsubsection{Second-order perturbation}

Next we consider the second-order perturbation, $\chi_2$.
The equation of motion is
\ba
\label{chi2-eq}
\ddot\chi_2+\left[3+(2\alpha-1)\kappa-\kappa\right]H\dot\chi_2
-\kappa\left[3+(2\alpha-1)\kappa\right]H^2\chi_2=S\, ,
\ea
where the source term, $S$, is given by
\ba
S=F_0\Biggl[\frac{(2-2\alpha)(1-2\alpha)}{2\kappa^2}
\left(\frac{\dot\chi_1}{H\phi_0}\right)^2
+\frac{(2-2\alpha)(2\alpha-1)}{\kappa}\frac{\dot\chi_1\,\chi_1}{H\phi_0^2}
+\frac{(2\alpha-1)(2\alpha-2)}{2}
\left(\frac{\chi_1}{\phi_0}\right)^2
\Biggr]\,.
\end{eqnarray}
Now we assume that the second solution of Eq.~(\ref{linearsol}),
$\chi_1\propto\exp[-(3+\eta-\kappa)Ht]$, dominates in $S$.
We find that $\chi_2\propto e^{\mu Ht}$ is a solution to
Eq.~(\ref{chi2-eq}), with  $\mu$ determined by the
time-dependence of $S$. We thus obtain
\ba
\phi =\phi_0+ \chi_1 + \left( \frac{g}{\phi_0} \right)  \chi_1^2 \, ,
\ea
with
\ba
g = \dfrac{(3 c_s^2+\kappa)}{4\kappa}(2-2\alpha) (1-2 \alpha)\, ,
\ea
and
\begin{eqnarray}
\mu=-2(3+\eta)+\kappa\, .
\end{eqnarray}

%%%%%%%%%%%%%%%%%%%%%%%%%%%%%%%%%%%%%%%%%%%%%%%%%%
\subsubsection{Calculating $\delta N$}

We are ready to compute the perturbations of the number of
$e$-folds, $\delta N$.
The background solution of $\phi$ (computed up to the
second-order perturbations around the reference trajectory, 
$\phi_0\propto e^{-\kappa N}$) in terms of $N$ is
\begin{eqnarray}
\nonumber
\phi &=&\phi_{0*}\left(e^{-\kappa N} +\lambda e^{(3+\eta-\kappa)N}
 +g \lambda^2 e^{(2(3+\eta)-\kappa)N}\right)
\\
&=& \frac{\phi_*}{1+\lambda+g\lambda^2}\,
\left(e^{-\kappa N} +\lambda e^{(3+\eta-\kappa)N}
 +g \lambda^2 e^{(2(3+\eta)-\kappa)N}\right)\,,
\label{phi-lambda2}
\end{eqnarray}
where $\lambda$ is an integration constant that parameterizes
different trajectories, and we have set $\phi(0,\lambda)=\phi_*$
for any value of $\lambda$ in accordance with the assumption that 
the end of the non-attractor phase is determined only by the value of the
scalar field, $\phi=\phi_*$.

Inverting Eq.~(\ref{phi-lambda2}) for a fixed $\lambda$, we would obtain
$N$ as a function of $\phi$ and $\lambda$. Then the $\delta N$ formula can be
obtained by 
\begin{eqnarray}
\delta N=N(\phi+\delta\phi,\lambda)-N(\phi,0)
=\sum_{n,m}\frac{1}{n!m!}
\frac{\partial^{n+m}N(\phi,0)}{\partial\phi^n\partial\lambda^m}
\delta\phi^n\lambda^m\,.
\end{eqnarray}
In practice, the explicit inversion of Eq.~(\ref{phi-lambda2}) is neither 
easy nor necessary. We may just assume $N$ in the right-hand side of it
as a function of $\phi$ and $\lambda$, $N=N(\phi,\lambda)$.
Then we may set $\phi\to\phi+\delta\phi$ and $N\to N+\delta N$
on both sides of Eq.~(\ref{phi-lambda2}) and solve for $\delta N$ iteratively.

So far, we have obtained approximate, perturbative
solutions of the scalar-field trajectories around the particular
reference solution, $\phi_0=\phi_*e^{\kappa Ht}$. These solutions are
valid only when the perturbed trajectories are not far away from the
reference solution, i.e., $|\chi_1+\chi_2|/\phi_0\ll 1$. However, the
dominant linear solution, $\chi_1/\phi_0=\lambda e^{-(3+\eta)Ht}\sim e^{3Ht}$,
quickly diverges as a function of time. Furthermore, as we can see from the
time-dependence of the second order solution, $\chi_2/\phi_0^2$ diverges even
faster, $\chi_2/\phi_0^2\sim e^{6Ht}$. From this, one suspects that
the approximate solutions can be trusted only for a short lapse of time.
In addition, since we have neglected the subdominant solution,
$\chi_1\propto e^{\kappa Ht}$, our approximation is valid only at sufficiently 
late times. These considerations suggest that we should choose the initial 
time as close as possible to the final time, $N\lesssim 1$. 
Then the simplest choice is
to take the initial time to be infinitesimally close to $t=t_*$.

Perturbing  $N=N(\phi,\lambda)$ up to the second order at $t=t_*$, we have
\begin{eqnarray}
\delta N = \left.\frac{\partial N}{\partial \phi}\right|_* \delta\phi_*
+ \left.\frac{\partial N}{\partial \lambda}\right|_* \lambda
+ \frac12\left.\frac{\partial^2 N}{\partial \phi^2}\right|_* \delta\phi_*^2
+ \frac12\left.\frac{\partial^2 N}{\partial \lambda^2}\right|_* \lambda^2
+ \left.\frac{\partial^2 N}{\partial \phi\partial\lambda}\right|_*
\delta\phi_*\lambda\, .
\end{eqnarray}
Using Eq.~(\ref{phi-lambda2}), we can easily evaluate the derivatives at 
the final hypersurface for a fixed $\phi_*$. In particular,
the $\lambda$-independence of $N$ at $N=0$ implies
${\partial N}/{\partial \lambda}|_*={\partial^2 N}/{\partial\lambda^2}|_*=0$.
Thus we need to evaluate only the other three terms, 
just as in Section~\ref{cs1}. This means that, in evaluating $\delta N$, we only need the linear terms in $\lambda$, while we have to take into account the $\phi$ dependence of $N$ up to the second order. That is, we can obtain $\delta N$, up to the second order, by using
\begin{eqnarray}
\phi=\frac{\phi_*}{1+\lambda}\,
\left(e^{-\kappa N} +\lambda e^{(3+\eta-\kappa)N}\right)\,.
\label{1stordersol}	
\end{eqnarray}
By taking the derivatives of both sides of Eq.~(\ref{1stordersol})
and setting $N=\lambda=0$ in the end, the necessary derivatives are 
easily computed as
\begin{eqnarray}
\left.\frac{\partial N}{\partial\phi}\right|_*
=\frac{1}{-\kappa\phi_{*}}\,,
\quad
\left.\frac{\partial^2 N}{\partial\phi\partial\lambda}\right|_*
=-\frac{3+\eta}{\kappa^2\phi_{*}}\,,
\quad
\frac12\left.\frac{\partial^2 N}{\partial\phi^2}\right|_*
=\frac{1}{2\kappa\phi_{*}^2}\,.
\label{derivatives}
\end{eqnarray}

Now we are to identify $\delta\phi_*$ and $\lambda$
with those generated from quantum fluctuations on flat slicing, $\delta\phi$.
To do so let us consider the evolution of $\delta\phi$ on superhorizon scales.
To the leading order in the slow-roll parameter $\epsilon$, which is an extremely
good approximation in the present case, $\delta\phi$ on flat slicing satisfies
exactly the same equation as $\chi$ for the background,
perturbatively given by Eqs.~(\ref{chi1-eq}) and (\ref{chi2-eq}).
Naturally $\delta\phi_1$ contains both growing and decaying modes initially,
where the subscript $1$ denotes it is of linear order.
 From Eq.~(\ref{linearsol}), we may set
\begin{eqnarray}
\delta\phi_1(N)=Ce^{(3+\eta-\kappa)(N-N_h)}+De^{-\kappa (N-N_h)}\,,
\end{eqnarray}
where $N_h$ is the number of $e$-folds at horizon crossing and one 
expects $|C|\sim|D|$. 
As the background trajectory is given by $\phi_0\propto e^{-\kappa N}$, 
it follows that the $D$-term, which has the same time-dependence as $\phi_0$,
corresponds to the adiabatic perturbation along the background trajectory.
Thus the $C$-term corresponds to the perturbation of the background trajectory.
Since the $D$-term is completely negligible at the end of the non-attractor
phase, $N=0$, we conclude
\begin{eqnarray}
\delta\phi_1(0)=\delta\phi_{1*}=\lambda\phi_*\,,
\label{dphistar}
\end{eqnarray}
hence to the second order,
\begin{eqnarray}
\delta\phi_*=\delta\phi_{1*}+\frac{g}{\phi_{*}}\delta\phi_{1*}^2\,,
\quad
\lambda=\frac{\delta\phi_{1*}}{\phi_{*}}\,.
\label{fluctuations}
\end{eqnarray}

Combining Eqs.~(\ref{derivatives}) and (\ref{fluctuations}),
we obtain $\delta N$ as
\ba
\delta N = -\dfrac{\delta\phi_{1*}}{\kappa \phi_*}
+ \left[-\kappa \left(g-\dfrac{1}{2}\right) -( \eta+3)\right]
\dfrac{\delta\phi_{1*}^2}{\kappa^2\phi_*^2}\, ,
\ea
from which we find $f_{NL}^{\rm local}$ as
\ba
\dfrac{3}{5} f_{NL}^{\mathrm{local}} 
= - \kappa g +\dfrac{\kappa}{2}  -(3 + \eta) 
&=& \dfrac{(3+\eta + 3 c_s^2)}{4 (1+c_s^2)} 
(1-\dfrac{1}{c_s^2}) -(3+\eta) + \dfrac{\eta c_s^2 }{2 (1+c_s^2)} \nonumber\\
&=& -\frac{1}{4 c_s^2} \left( 9 c_s^2 + 2 \eta c_s^2 + 3 + \eta \right) \, .
\ea
With  $\eta=-6$ to obtain a scale-invariant power spectrum,
\ba
 f_{NL}^{\mathrm{local}} = \dfrac{5}{4 c_s^2} (1+c_s^2) \, .
\label{fNL-deltaN}
\ea
This result is valid for any values of $c_s$, as
we have not assumed $c_s\ll 1$, and agrees exactly with
 the result obtained from the in-in formalism
 (Eq.~\eqref{fNL-comoving}). Furthermore, we obtain $
 f_{NL}^{\mathrm{local}} = 5/2$ for $c_s=1$, in agreement with the
 result we find in Section~\ref{cs1}.

\section{Conclusion}
\label{conclusion}

The non-attractor inflation models giving ${\cal R}\propto a^3$ on
super-horizon scales are so
far the only examples of self-consistent single-field inflation models
based on a Bunch-Davies initial state
that give a scale-invariant power spectrum and a large squeezed-limit
bispectrum, violating Maldacena's consistency relation \cite{Namjoo:2012aa,
Chen:2013aj,Martin:2012pe,Huang:2013lda}. The previous work \cite{Namjoo:2012aa,
Chen:2013aj} shows that the local-form bispectrum parameters from these
models are $f_{NL}^{\rm local}=5/2$ and $3/(4c_s^2)$ for $c_s=1$ and
$c_s\ll 1$, respectively. Therefore, detection of a large local-form
bispectrum violating Maldacena's consistency relation would not rule out
all single-field inflation models in full generality; rather, it would rule out
all single-field inflation models which are based on a Bunch-Davies
initial state and have reached the attractor solutions.

Given the importance of this statement, in this paper we have provided
more detailed derivation. We find that two completely
different methods, the quantum field
theory calculation using the in-in formalism in the comoving gauge and
the classical calculation using the $\delta N$ formalism, give the same
result, $f_{NL}^{\rm local}=5(1+c_s^2)/(4c_s^2)$, which is valid for
arbitrary values of $c_s$. This is because the non-attractor model
generates non-Gaussianity on super-horizon scales, as the interactions
such as $\dot{\cal R}^3$ and ${\cal R}\dot{\cal R}^2$, which yield the
equilateral bispectrum in the usual attractor case (for which $\dot{\cal
R}\approx 0$ on super-horizon scales), become proportional to ${\cal R}^3$ on
super-horizon scales via $\dot{\cal R}=3H{\cal R}$. We also find that
the third method using the in-in formalism in the flat gauge and
decoupling limit ($c_s\ll 1$) gives the same answer in the appropriate
limit. In contrast to the usual attractor single field case, this model does not predict observable equilateral bispectrum.

While (a shorter version of) the derivation of $f_{NL}^{\rm
local}=3/(4c_s^2)$ for $c_s\ll 1$ using the in-in formalism has already
been presented in Ref.~\cite{Chen:2013aj}, the full
derivation for arbitrary $c_s$ and the derivation using the
$\delta N$ formalism are new.
As the scalar field trajectory is determined by two parameters, we
usually specify the scalar field value and its derivative at some epoch,
$\phi_i$ and $\dot\phi_i$, when applying the $\delta N$ formalism
\cite{Namjoo:2012aa}. If the solution has reached the attractor
solution, $\dot\phi_i$ is uniquely determined by $\phi_i$, and thus we
need to differentiate the number of $e$-folds with respect to $\phi_i$ only;
otherwise, we must differentiate the number of $e$-folds with respect to
both $\phi_i$ and $\dot\phi_i$. In this paper, we find it more
convenient to use two parameters naturally characterizing the
scalar field trajectories, which are not necessarily $\phi_i$ or
$\dot\phi_i$. We have used this methodology to derive both $f_{NL}^{\rm
local}=5/2$ for $c_s=1$ (which was previously derived by
Ref.~\cite{Namjoo:2012aa} using $\phi_i$ and $\dot\phi_i$)  and
$f_{NL}^{\rm local}=5(1+c_s^2)/(4c_s^2)$ for $c_s\ne 1$
(which had not previously been derived).

%%%%%%%%%%%%%%%%%%%%%%%%%%%%%%%%%%%%%%%%%%%%%%%%%

 \section*{Acknowledgment}
 We would like to thank A. A. Abolhasani and A. Naruko for useful discussions. 
XC is supported in part by the Stephen Hawking Advanced Fellowship.
This work was supported in part by 
JSPS Grant-in-Aid for Scientific Research (A) No.~21244033.

 %%%%%%%%%%%%%%%%%%%%%%%%%%%%%%%%%%%%%%%%%%%%%%%%%

\appendix

\section{Equivalence of Lagrangians in the comoving and flat gauges}
\label{App:Equiv}

In this Appendix, we prove that two Lagrangians in the comoving
(Section~\ref{Sec:comoving}) and flat (Section~\ref{Sec:flat}) gauges are
equivalent to each other at the leading order in $c_s^{-2}$ and
$\lambda/\Sigma$. For the comoving gauge, the leading terms
include the first three terms in Eq.~(\ref{action3}) and the
field-redefinition terms in Eq.~(\ref{field_redef}); for the flat gauge,
they include both terms in Eq.~(\ref{action_deltaphi2}).
We assume that $H$, $\dot\phi$, $\eta$, and $c_s^2$ are approximately
constant, and $c_s^{-2}$ and  $\lambda/\Sigma$ are much bigger than unity.

The difference between the first three terms in Eq.~(\ref{action3}) and
both terms in Eq.~(\ref{action_deltaphi2}) is
\bea
\CL_3^\calR - \CL_3^{\delta\phi}
= \frac{\epsilon}{c_s^4} a^3 \frac{\dot\calR^3}{H}
- \frac{3\epsilon}{c_s^4} a^3 \calR \dot\calR^2
+ \frac{\epsilon}{c_s^2} a \calR (\partial \calR)^2
- \frac{\epsilon}{c_s^2} \frac{a}{H} \dot\calR (\partial\calR)^2 ~.
\label{difference}
\eea
Defining
\bea
\frac{-1}{2a^3} \frac{\delta\CL_2}{\delta\calR}
\equiv
\frac{\epsilon}{c_s^2} \ddot\calR
+ (3+\eta) H \frac{\epsilon}{c_s^2} \dot\calR
- \frac{\epsilon}{a^2} \partial^2\calR ~,
\label{dL2dphi}
\eea
and integrating by parts the first term in Eq.~(\ref{difference}), Eq.~(\ref{difference}) becomes
\bea
\frac{1}{H c_s^4}
\left(
\eta\epsilon H a^3 \dot\calR^2 \calR
- 2a \epsilon c_s^2 \calR \dot\calR \partial^2 \calR
+ \epsilon c_s^2 a H \calR (\partial\calR)^2
- \epsilon c_s^2 a \dot\calR (\partial\calR)^2
\right)
+ \frac{1}{H c_s^2} \calR \dot\calR \frac{\delta\CL_2}{\delta\calR}
~,
\eea
where we have dropped a temporal total derivative proportional to
$\frac{d}{dt} ( \epsilon a^3 \dot\calR^2 \calR )$.

Using the following integration by parts (ignoring spatial total derivatives),
\bea
a \calR \dot\calR \partial^2 \calR \to
\frac{1}{4} \frac{d}{dt} (a \calR^2 \partial^2\calR)
+ \frac{a H}{2} \calR (\partial\calR)^2
- \frac{a}{2} \dot\calR (\partial\calR)^2 ~,
\eea
the difference between Lagrangians in two gauges given in
Eq.~(\ref{difference}) becomes
\bea
\frac{1}{H c_s^4}
\left(
\eta \epsilon H a^3 \dot\calR^2 \calR
+ \half \eta \epsilon H c_s^2 a \calR^2 \partial^2\calR
\right)
+ \frac{1}{H c_s^2} \calR \dot\calR \frac{\delta\CL_2}{\delta\calR}
~,
\label{difference2}
\eea
where we have dropped a temporal total derivative proportional to
$\frac{d}{dt} ( \epsilon a \calR^2 \partial^2 \calR )$.
Integrating by parts the first term in Eq.~(\ref{difference2}) and
ignoring $\dot\eta$, this term gives
\bea
-\frac{1}{2} \eta\epsilon H a^3 \calR \ddot\calR
-\frac{1}{2} \epsilon\eta (\eta+3) H^2 a^3 \dot\calR \calR^2 ~,
\label{first_term}
\eea
where we have dropped a temporal total derivative proportional to
$\frac{d}{dt} ( \epsilon a^3 \dot\calR \calR^2 )$.

Inserting Eq.~(\ref{first_term}) into Eq.~(\ref{difference2}), the
difference finally becomes
\bea
\left(
 \frac{\eta}{4 c_s^2} \calR^2 + \frac{1}{c_s^2 H} \dot\calR \calR
\right)
\frac{\delta\CL_2}{\delta\calR} ~,
\eea
which is equivalent to the field-redefinition terms given in
Eq.~(\ref{field_redef}).
The three temporal total-derivative terms we have dropped give no
contribution to the three-point function, as the contraction of two
$\calR$'s at the equal time vanishes. This ends the proof.

%%%%%%%%%%%%%%%%%%%%%%%%%%%%%%%%%%%%%%%%%%%%%%%%%%%%%%%%%%%%
{}

\end{document}